\documentclass[reprint,amsmath,amssymb,aps,prl]{revtex4-2}

\usepackage{graphicx} 
\usepackage{bm} 
\usepackage{amsmath,amssymb}
\usepackage{xcolor}
\usepackage{array}
\usepackage{tabularx}

\graphicspath{ {./}{Figures/}}
\begin{document}

\title{\textit{Ab initio} prediction of strain-tunable spin defects in quasi-1D TiS$_3$ and NbS$_3$ nanowires}

\author{Jordan Chapman$^{1}$}
\author{Arindom Nag$^{2}$}
\author{Thang Pham$^{2}$}
\author{Vsevolod Ivanov$^{1,3,4}$ }
\email{vivanov@vt.edu}

\affiliation{$^1$Virginia Tech National Security Institute, Blacksburg, Virginia 24060, USA}
\affiliation{$^2$Department of Materials Science and Engineering, Virginia Tech, Blacksburg, Virginia 24061, USA}
\affiliation{$^3$Department of Physics, Virginia Tech, Blacksburg, Virginia 24061, USA}
\affiliation{$^4$Virginia Tech Center for Quantum Information Science and Engineering, Blacksburg, Virginia 24061, USA}

\begin{abstract}
Defects in atomically thin van der Waals materials have recently been investigated as sources of spin-photon entanglement with sensitivity to strain tuning. Unlike many two-dimensional materials, quasi-one-dimensional materials such as transition metal trichalcogenides exhibit in-plane anisotropy resulting in axis-dependent responses to compressive and tensile strains. Herein, we characterize the tunable spin and optical properties of intrinsic vacancy defects in titanium trisulfide (TiS$_3$) and niobium trisulfide (NbS$_3$) nanowires. Within our \textit{ab initio} approach, we show that sulfur vacancies and divacancies (V$_S$ and V$_D$, respectively) in TiS$_3$ and NbS$_3$ adopt strain-dependent defect geometries between in-plane strains of -3 \% and 3 \%. The calculated electronic structures indicate that both V$_S$ and V$_D$ possess in-gap defect states with optically bright electronic transitions whose position relative to the conduction and valence bands varies with in-plane strain. Further, our calculations predict that V$_S$ in TiS$_3$ and V$_D$ in NbS$_3$ exhibit transitions in their ground state spins; specifically, a compressive strain of 0.4 \% along the direction of nanowire growth causes a shift from a triplet state to a singlet state for the V$_S$ defect in TiS$_3$, whereas a tensile strain of 2.9 \% along the same direction in NbS$_3$ induces a triplet ground state with a zero-phonon line of 0.83 eV in the V$_D$ defect. Our work shows that the anisotropic geometry of TiS$_3$ and NbS$_3$ nanowires offers exceptional tunability of optically active spin defects that can be used in quantum applications.

\end{abstract}

\maketitle

\section{Introduction}

Color center defects in semiconductors are an attractive platform for applications in quantum information science \cite{Wasielewski2020,Bayliss2020}, as they couple light and electronic spin states, enabling scalable spin-photon qubit architectures \cite{Landig2018}. Several promising defect candidates for spin qubits and single photon emitters (SPEs) have been realized in a number of bulk semiconductors including diamond, silicon carbide, and silicon \cite{awschalom2010_pnas, Burkard2023, Redjem2020}, and more recently in two-dimensional (2D) materials \cite{Zeng2024, Guo2024}. 

Strain engineering of color centers in semiconductor materials has been investigated as a strategy to tune the emission properties of atomic defects \cite{Meesala2018,Ristori2023,Brevoord2025,Tissot2024}. Low-dimension van der Waals (vdW) materials boast several distinct advantages over bulk materials for strain engineering of color center defects thanks to their atomically thin nature. Nanoscale vdW materials have been shown to exhibit drastically enhanced mechanical flexibility \cite{Kim2009} compared to bulk counterparts, and their strains can be modulated via numerous methods, including attachment to an atomic force microscopy (AFM) cantilever \cite{Bertolazzi2011}, heating of gas molecules between layers \cite{Tyurina2019}, and bending of a substrate material \cite{Conley2013}. Many 2D vdW materials, notably graphene, hexagonal boron nitride (hBN), and transition metal dichalcogenides (TMDCs), have been explored as hosts for strain-tunable quantum point defects \cite{Santra2024}.
For instance, strained defect SPEs within a single layer of tungsten diselenide \cite{Parto2021} were found to exhibit narrow zero-phonon lines (ZPLs) of about 775 nm with widths measured at 75 $\mu$eV, whereas single photon properties were not observed for unstrained defect structures.
Density functional theory (DFT) calculations revealed that the observed emission peaks could not be attributed to a single defect (Se vacancy, W vacancy, or O interstitial \cite{Komsa2013}) and that the bright emission is likely due to the formation of a WSe$_6$ pore vacancy. Negatively charged boron vacancy (V$^-_\text{B}$) spin defects in hBN \cite{Stern2022} are sensitive to strain \cite{hBN_review}, which has led to their use in quantum strain sensing applications \cite{Aharonovich2022, Lyu2022}.

\begin{figure*}[htb]
    \includegraphics[width=1.0\textwidth,trim=8 0 0 0,clip]{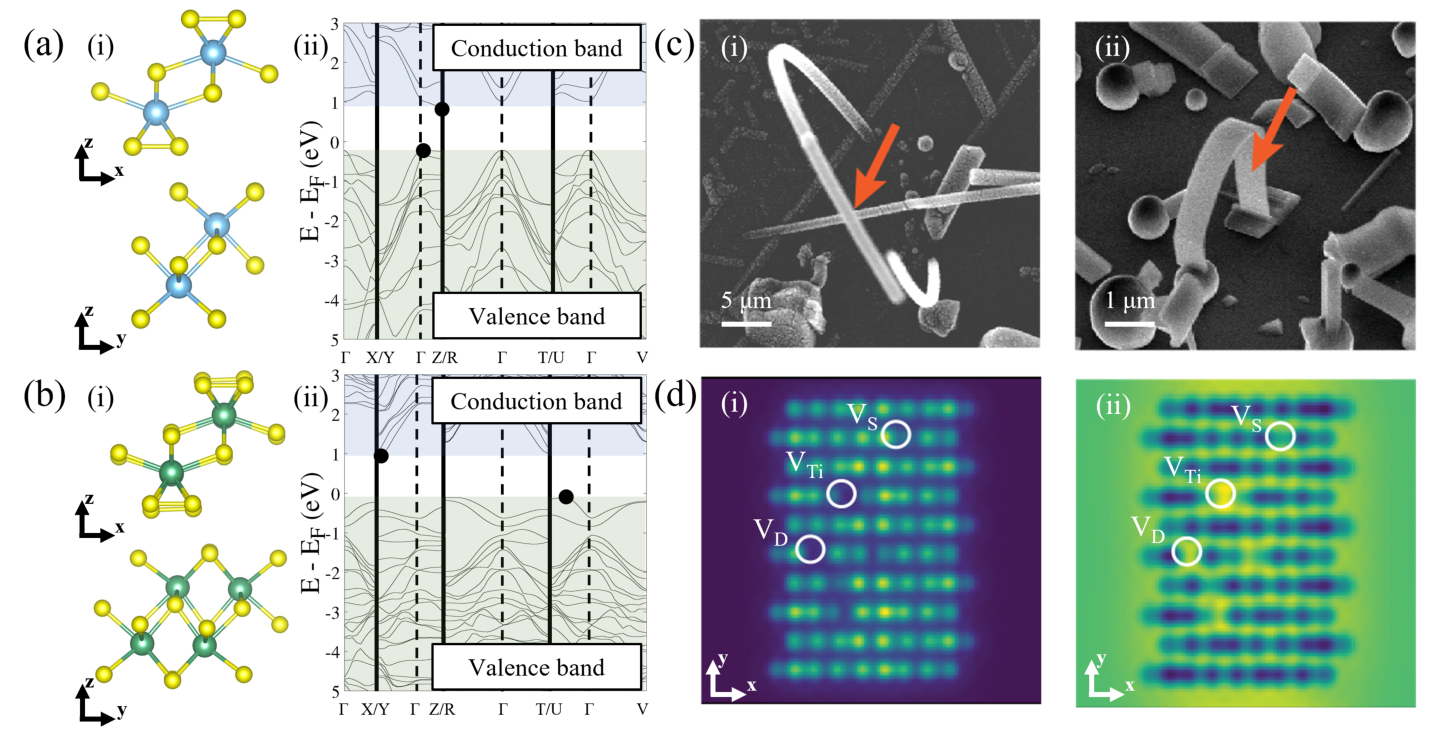}
    \caption{Theoretical predictions of TMTC unit cell geometry (i) and electronic band structure (ii) for (a) TiS$_3$ and (b) NbS$_3$. Band energies are plotted relative to the Fermi energy. (c) SEM images of bent (i) TiS$_3$ and (ii) NbS$_3$ nanowires (indicated by an orange arrow). (d) Simulated scanning transmission electron microscopy (STEM) images showing common intrinsic defects (V$_S$: sulfur vacancy, V$_D$: sulfur divacancy and V$_{Ti}$: titanium vacancy) in TiS$_3$ nanowires in dark field (i) and bright field (ii) modes. }  
    \label{Pristine}
\end{figure*}

A family of materials related to TMDCs are transition metal trichalcogenides (TMTCs), which form quasi-one-dimensional (1D) MX$_3$ nanowires (M = Ti, Hf, or Nb and X = S, Se, or Te). TMTCs exhibit a wide range of structure-dependent transport properties, inspiring their application as high-pressure superconductors \cite{Yue2023}, charge density wave (CDW) materials \cite{Semakin2024,Zybtsev2017}, and visible and infrared emitters \cite{Khatibi2020,Conejeros2021}. Unlike many 2D materials, TMTCs have in-plane anisotropy, which leads to axis-dependent strain responses \cite{Li2015}. The electronic band structures of TMTCs are quite resistant to stacking order when compared to those of 2D materials; this has been attributed to the electronic states near the band edge being localized to the internal transition metal-sulfur (M-S) bonding direction \cite{Kang2016}. 

The anisotropic relationships among bulk TMTC properties are well studied \cite{Kang2016,Saiz2020,Shu2022,Zhang2024,Gan2019,Saeed2017}, but exploration of their intrinsic defect properties has been comparatively limited in scope \cite{Tian2020,Formo2022,Zheng2016,Iyikanat2015,Khatibi2020}. Recent DFT efforts to systematically characterize intrinsic defects in TiS$_3$, including S and Ti vacancies up to relatively high doping concentrations, revealed that these defects can contribute to \textit{n}- or \textit{p}-type conductivity and populate deep in-gap states \cite{DonghoNguimdo2024}. The effect of S vacancies on exciton lifetimes in TiS$_3$ was also investigated to explain discrepancies in electron-hole recombination rates between experimental and theoretical studies \cite{Wei2017}. Pairing time domain \textit{ab initio} calculations and MD, it was found that S vacancies reduce the non-radiative recombination rate of excitons in TiS$_3$ by an order of magnitude. However, the accuracy of these efforts may be limited due to their use of PBE functionals \cite{GGA}, which are known to underestimate electronic band gaps and defect localization \cite{Borlido2020_bandgap_problem, gali_hse_deltascf}.

Inspired by previous reports on strain engineering of defect-induced SPEs in 2D vdW materials and our recent synthesis of TiS$_3$ and NbS$_3$ nanowires (Figures \ref{Pristine}(c,d)), we use first principles calculations to characterize the local structure of intrinsic chalcogen vacancies in TiS3$_3$ and NbS$_3$ in response to uniaxial strain. We identify single-photon emitters with ZPLs ranging from 0.3eV (mid-IR) to 0.6eV (near-IR) with strain tunability of their ZPLs up to $\pm$ 0.24 eV. Additionally, we find that strain can also be used to induce a transition between singlet and triplet ground states in the sulfur vacancy (V$_S$) defect in TiS$_3$ and the sulfur divacancy (V$_D$) defect in NbS$_3$. The inherent anisotropy and malleable optoelectronic properties of TMTC nanowires make them strong candidate materials for hosting entangled spin-photon defects with the additional advantage of strain-driven spin state transitions.

\section{Methods}
 
\noindent \textbf{\textit{Ab initio} calculations.} All calculations were performed with the Vienna Ab Initio Simulation Package (VASP) \cite{VASP1,VASP2}, which uses plane-wave basis sets following the projector augmented wave (PAW) method \cite{Blochl2994}. All geometry relaxations and electronic self-consistency calculations were converged to 10$^{-3}$ eV Å$^{-1}$ and 10$^{-10}$ eV, respectively. Sampling of the Brillouin zone for all structures was done at the $\Gamma$ point with a cutoff energy of 650 eV. 

TMTC supercells were constructed from respective unit cells (shown in Figures \ref{Pristine}(a,b)) to host intrinsic S vacancy defects; the TiS$_3$ unit cell was extended to a $4 \times 4 \times 1$ supercell (128 atoms), while the NbS$_3$ unit cell was extended to a $3 \times 3 \times 1$ supercell (144 atoms). A vacuum layer of 15 \AA{} was added in the $z$-direction to simulate TMTC monolayers. Pristine crystal supercells were geometry- and volume-relaxed using a screened hybrid functional. Contributions from DFT were calculated using a generalized gradient approximation \cite{GGA} including vdW dispersion energy corrections calculated according to the zero-damping DFT-D3 method \cite{Grimme2010}. We find that the Heyd-Scuseria-Ernzerhof (HSE06) functional \cite{HSE} well reproduces the electronic band gaps of both pristine TiS$_3$ and NbS$_3$ crystal structures. Volume relaxation results in good agreement with experimental lattice parameters for TiS$_3$ \cite{Dai2015} but an overestimation of 1.1 \% of the \textit{b}-lattice parameter for NbS$_3$ \cite{Bloodgood2017}. Because the band edges and defect levels are sensitive to small amounts of strain, we expect that straining the crystal structures relative to their volume-relaxed geometries will provide a better description of strain-dependent electronic, optical, and spin properties. Strains ranging from -3 \% (compressive) to 3 \% (tensile) were induced independently in both in-plane directions (denoted by $\varepsilon_x$ and $\varepsilon_y$) by adjusting the \textit{a}-lattice and \textit{b}-lattice parameters followed by geometry relaxation. 

\begin{figure}[t]
    \includegraphics[width=1.0\columnwidth]{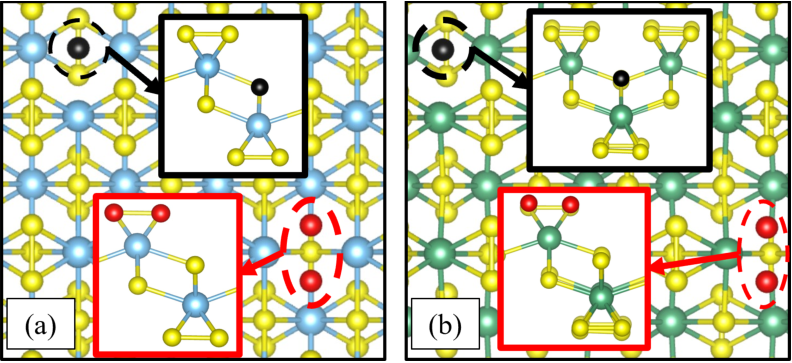}
    \caption{Vacancy defects in (a) TiS$_3$ and (b) NbS$_3$. V$_S$ and V$_D$ defects are shown in black and red, respectively.}
    \label{Defects_visual}
\end{figure}

The anisotropy of TMTC crystals gives rise to chemically distinct S ion species within the unit cells of TiS$_3$ and NbS$_3$. Single V$_S$ and V$_D$ defects were embedded within the TMTC supercell structures to elucidate strain-dependent defect properties. We denote V$_S$ as a vacancy of an internal S atom and V$_D$ as a pair of nearest-neighbor S vacancies that make up the short edge of the triangular prismatic polyhedral surface, as shown in Figure \ref{Defects_visual}. Defect-embedded supercells were geometry relaxed and subjected to the same lattice straining method performed for pristine TMTC supercells. Formation energies of V$_S$ and V$_D$ defects were calculated according to:
\begin{equation*}
    E_\text{f}(\varepsilon_j) = E_\text{defect}(\varepsilon_j) - E_\text{pristine}(\varepsilon_j) + \Sigma\mu_\text{i}n_\text{i}~,
\end{equation*}
where $E_\text{f}$ is the formation energy, $E_\text{defect}$ is the total energy of the defective system, $E_\text{pristine}$ is the total energy of the pristine crystal, and $\varepsilon_j$ (j= x, y) is the axis-dependent strain. Changes in the chemical potential are accounted for in the last term, where $\mu$$_i$ and n$_i$ are the chemical potential and number of the removed S ions, respectively. We also probe the electronic structures corresponding to the simple excitations of the in-gap defect states via the constrained occupation method \cite{Gali2009}; optical properties of the corresponding excitations were calculated using VASPKIT \cite{VASPKIT}.

\begin{figure}[b]
    \includegraphics[width=1.0\columnwidth,trim=8 0 0 0,clip]{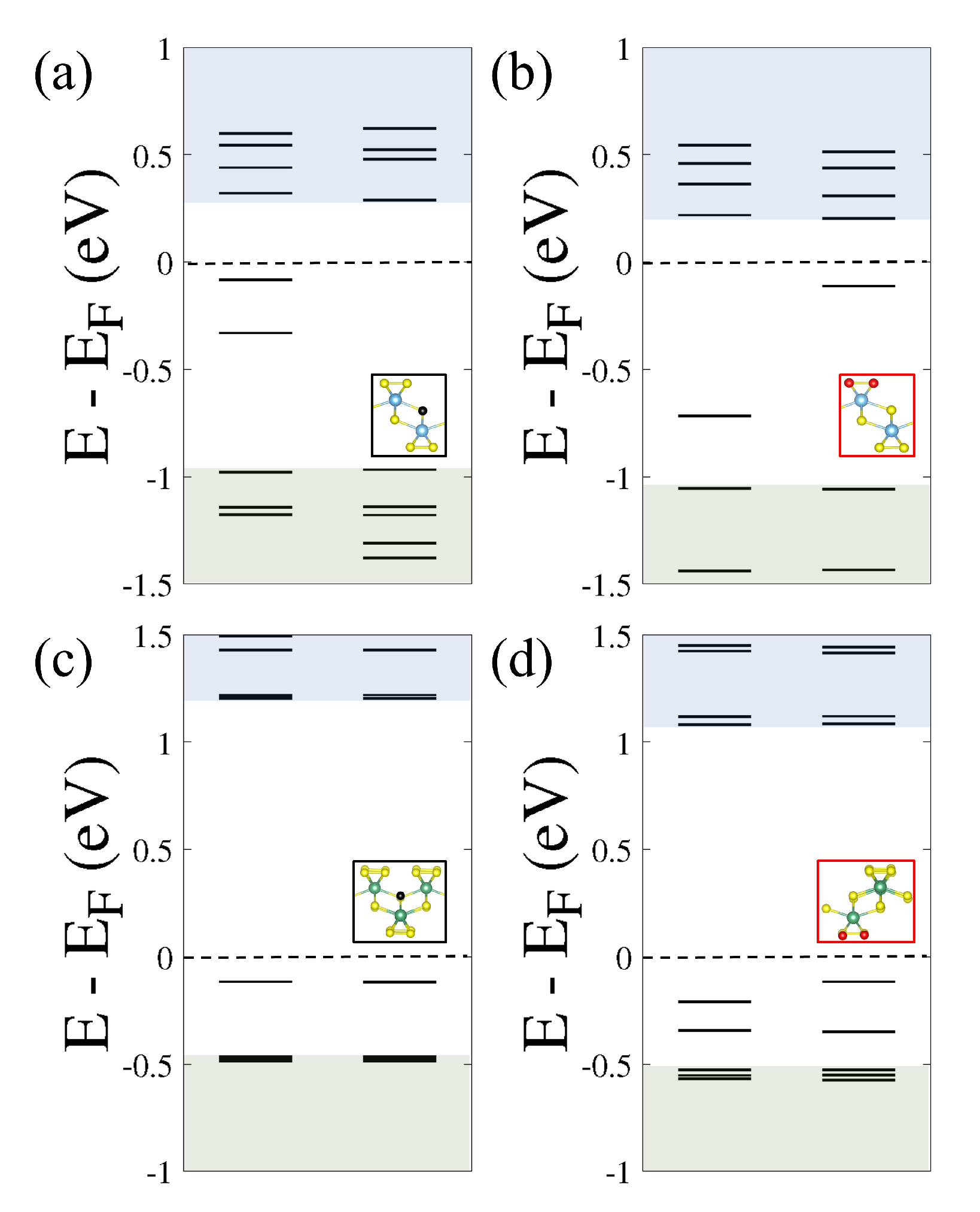}
    \caption{Electronic structures of (a) V$_S$, (b) V$_D$ in a $4 \times 4 \times 1$ TiS$_3$ supercell, (c) V$_S$, (d) V$_D$ in a $3 \times 3 \times 1$ NbS$_3$ supercell, showing the conduction band (blue), valence band (green), localized defect states (black lines) plotted relative to the Fermi energy $E_F$ (dashed line). Up/down spin channels are shown separately in the left/right columns of each panel. 
    }
    \label{e-_structures}
\end{figure}

\noindent \textbf{Synthesis and characterization of transition metal trichalcogenide nanowires.} TiS$_3$ and NbS$_3$ nanowires were synthesized by salt-assisted chemical vapor deposition. Details of the experimental setup and the synthesis parameters can be found in previously published work \cite{Pham2024}. Scanning Electron Microscopy (SEM) images were acquired by Helios 5 UC with 5-10 keV beam energy. Multi-slice simulations of scanning transmission electron microscopy (STEM) images were conducted using abTEM package \cite{abTEM} with the microscope conditions similar to the experimental parameters (electron energy of 200 keV and convergence angle of 25 mrad).

\section{Results and Discussion}

\noindent \textbf{Strain tuning of pristine nanowires.} Our \textit{ab initio} calculations reveal that pristine TiS$_3$ and NbS$_3$ nanowires have axis-dependent responses to compressive and tensile strains. Electronic band structures of unstrained TiS$_3$ and NbS$_3$ are plotted in Figures \ref{Pristine}(a,b). While the TiS$_3$ band gap of 1.10 eV is in good agreement with literature \cite{Dai2015,Kang2016}, we find that its band gap is indirect rather than direct. We predict the energy difference between the direct and indirect band gap to be only 0.016 eV, so it is likely that this discrepancy is due to slight symmetry breaking of the pristine TiS$_3$ crystal in our approach. We also reproduce the 1.22-eV indirect band gap of NbS$_3$  \cite{Conejeros2021}.

Observation of many bent nanowires, both TiS$_3$ and NbS$_3$ (Figure \ref{Pristine}(c)) in our growth indicates their mechanical flexibility. We next perform calculations on strain-response of the supercell structures and the electronic band structure. TiS$_3$ and NbS$_3$ supercells exhibit similar trends in response to anisotropic axial strains, both of which are plotted in Figure S.1 (see Supplementary Materials \cite{supplement}). Compressive strains incur larger increases in the ground state energies of the pristine supercells, suggesting that the nanowire structures are more tolerant of tensile strains. These results are consistent with findings previously reported for 2D vdW materials, which can tolerate only small compressive strains before buckling out-of-plane \cite{Santra2024}. Further, both compressive and tensile strain along the $y$-direction results in sharper increases in the ground state energies when compared to strain along the $x$-direction for both TMTC nanowires. The enhanced sensitivity to $\varepsilon_y$ can be understood as a consequence of the strong covalent nature of M-S bonding in the y-direction, while M-S bonds in the $x$-direction of both TiS$_3$ and NbS$_3$ are comprised of weak covalent bonds and vdW interactions.

Similarly, TiS$_3$ and NbS$_3$ band gaps are more strongly dependent on $\varepsilon_y$ than $\varepsilon_x$. Band gap responses to external strain (Figure S.1) are in good agreement with literature values \cite{Conejeros2021,Saiz2020}. External strain in the y-direction results in a near-linear modulation of the band gap. We find that depending on the direction of strain, the band gaps of TiS$_3$ and NbS$_3$ can be increased or decreased by about 21 \%  and 15 \% respectively, relative to the unstrained values. Wavefunction visualization of electronic states near the band edges (Figure S.2) confirms that M-S bonds are well hybridized along the y-direction in each TMTC crystal \cite{Kang2016}. Band structure calculations from $\varepsilon_y$ = -3 \% to 3 \% and $\varepsilon_x$ = -3 \% to 3 \% are plotted in Figure S.3 \cite{supplement}; we find that that the positions of the valence band maximum (VBM) and conduction band minimum (CBM) are not sensitive to in-plane strain. 

\noindent \textbf{Optically active sulfur vacancies.} Transition metal chalcogenides are prone to chalcogen (e.g., S) vacancies because of their high vapor pressure during the material growth. Our simulated scanning transmission electron microscopy (STEM) images (Figure \ref{Pristine}(d)) show common intrinsic defects in TiS$_3$ nanowires, including V$_S$, V$_D$ and V$_{Ti}$, which can be ambiguously identified. We perform calculations on  the electronic, optical, and spin properties of selected intrinsic vacancy defects in quasi-1D TiS$_3$ and NbS$_3$ crystals using a hybrid-functional approach. We consider only V$_S$ and V$_D$ defects (Figure \ref{Defects_visual}) in the dilute limit to mitigate the effects of self-interactions with periodic defect images that can occur for larger defect geometries and delocalized defect states.

\begin{table}[hb]
    \caption{Defect properties of intrinsic S vacancy defects in TiS$_3$ and NbS$_3$. We report the optical transitions corresponding to the two highest in-gap states of each defect. Formation energy values are compared to those reported by Iyikanat \textit{et al} \cite{Iyikanat2015}.}
    \vspace{0.5cm}
    {\renewcommand{\arraystretch}{1.3}
    \begin{tabular}{c | c c c c c c c c c c c}
        \hline
        \hline
        Host   &   Defect   & &   $E_{(F)}^\text{th}$   & &   $E_{(F)}^\text{Lit}$   & &     \textbf{S}   & &   E$_{ZPL}$   & &   TDM$^2$    \\
         material               &     & &   (eV)   & &   (eV)   & &        & &   (eV)   & &    (Debye$^2$)   \\
        \hline
        TiS$_3$ & V$_S$ && 1.08 && 3.58 && 1 && 0.26 && 50.3 \\
                &       &&      &&      &&   && 0.31 && 18.2 \\
                & V$_D$ && 1.62 && 8.48 && 0 && 0.23 && 2.10 \\
                &       &&      &&      &&   && 0.23 && 6.02 \\
        \hline
        NbS$_3$ & V$_S$ && 1.52 && -    && 0 && 0.61 && 0.12 \\
                &       &&      &&      &&   && 0.62 && 16.8 \\
                & V$_D$ && 2.02 && -    && 0 && 0.60 && 0.60 \\
                &       &&      &&      &&   && 0.65 && 0.41 \\
        \hline
        \hline
    \end{tabular}
    }
    \label{defectdata}
\end{table}

\begin{figure*}[ht!]
    \includegraphics[width=1.0\textwidth,trim=15 20 0 5,clip]{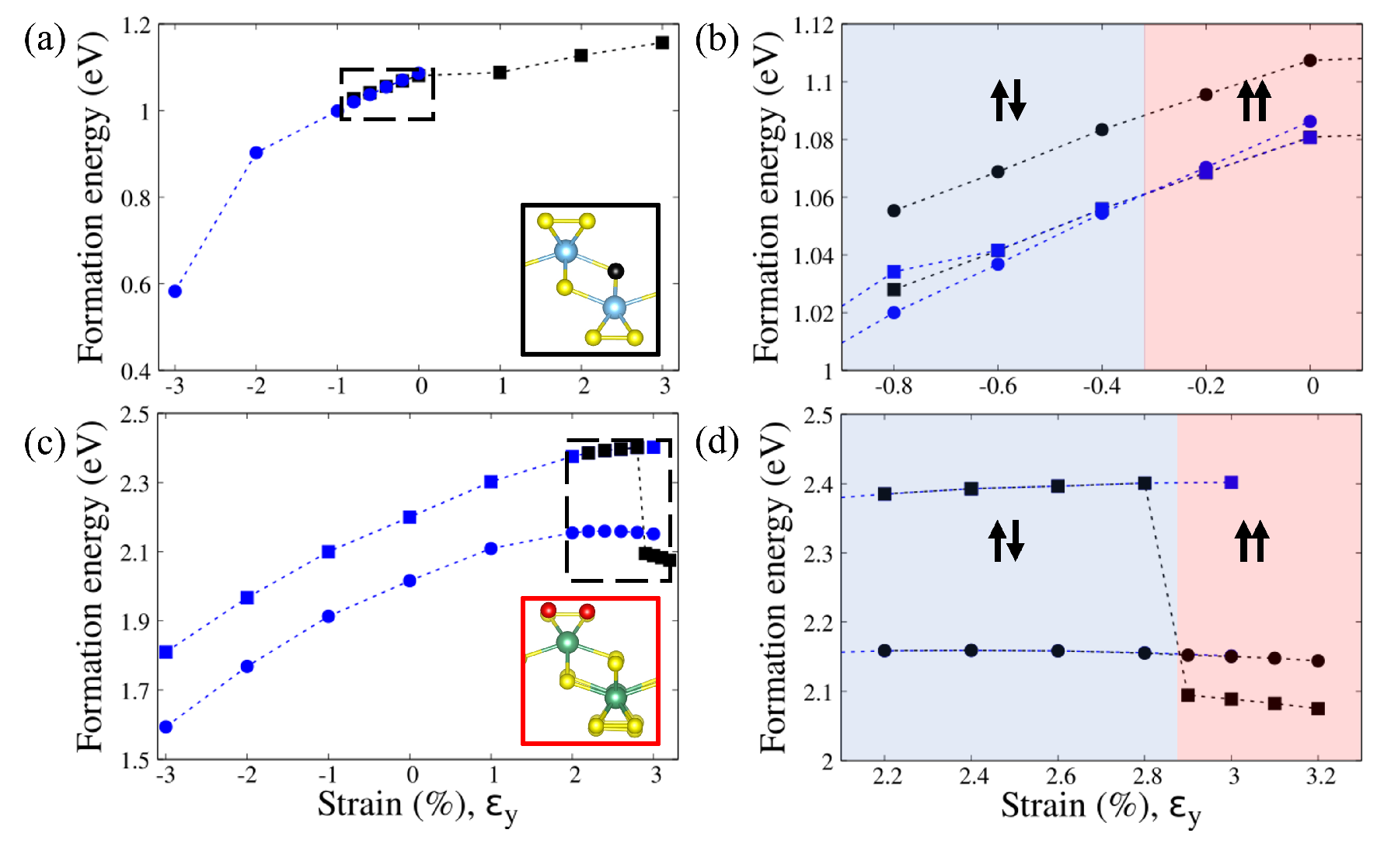}
    \caption{Strain response of formation energies of intrinsic vacancy defects. Formation energy as a function of $\varepsilon_y$ and the same curve with increased resolution are shown in (a) and (b), respectively, for V$_S$ in TiS$_3$ and (c) and (d) for V$_D$ in NbS$_3$. Singlet ground states are plotted with circles, while square marks correspond to triplet ground states. Defect structural polymorphs are delineated with different color curves.}
    \label{defect_strain_formation}
\end{figure*}

The electronic structures of single V$_S$ and V$_D$ defects in TiS$_3$ and NbS$_3$ supercells are shown in Figure \ref{e-_structures}. Our calculations reveal that both V$_S$ and V$_D$ defects possess in-gap states with optically bright electronic transitions. The computed defect properties, including formation energies, ZPLs, transition dipole moments (TDMs), and spin state are summarized in Table \ref{defectdata}. We find that incorporation of S vacancy defects slightly widens the electronic band gaps of both TiS$_3$ and NbS$_3$, consistent with previously reported theoretical studies of S vacancies in TiS$_3$ \cite{Wei2017,Iyikanat2015}. This behavior is attributed to decreased hybridization between transition metal and S orbitals, which in turn pushes the highest occupied molecular orbital (HOMO) level downward.

We can compare the formation energies for S vacancies in TiS$_3$ with previously reported calculations \cite{Iyikanat2015,DonghoNguimdo2024}. Our calculations predict defect formation energies in TiS$_3$ that are smaller than those reported by Iyikanat \textit{et al}, likely in part due to calculation of ground state energies at different levels of theory and with different supercell sizes. We calculated defect energies using hybrid DFT and in a 4 x 4 x 1 supercell, while the previous work uses a smaller 3 x 3 x 1 supercell at the PBE level of theory, both of which may explain the predicted metallic behavior of their V$_D$-doped TiS$_3$ system \cite{Rana2022}.

Vacancy defect levels lie higher in the band gap for TiS$_3$ than for NbS$_3$. V$_S$ and V$_D$ defect states in TiS$_3$ are somewhat hybridized with electronic states that populate the CBM, whereas analogous defects in NbS$_3$ are hybridized with states that comprise the VBM. While TiS$_3$ and NbS$_3$ have similar band gaps, the optical transitions of intrinsic defects in NbS$_3$ are associated with larger ZPL energies and thus emit closer to the telecom band. These emission properties are congruent with previous work detailing the near-IR emission of bulk TiS$_3$ and NbS$_3$ materials \cite{Khatibi2020}.

\begin{figure*}[ht]
    \includegraphics[width=1.0\textwidth,trim=10 12 0 0,clip]{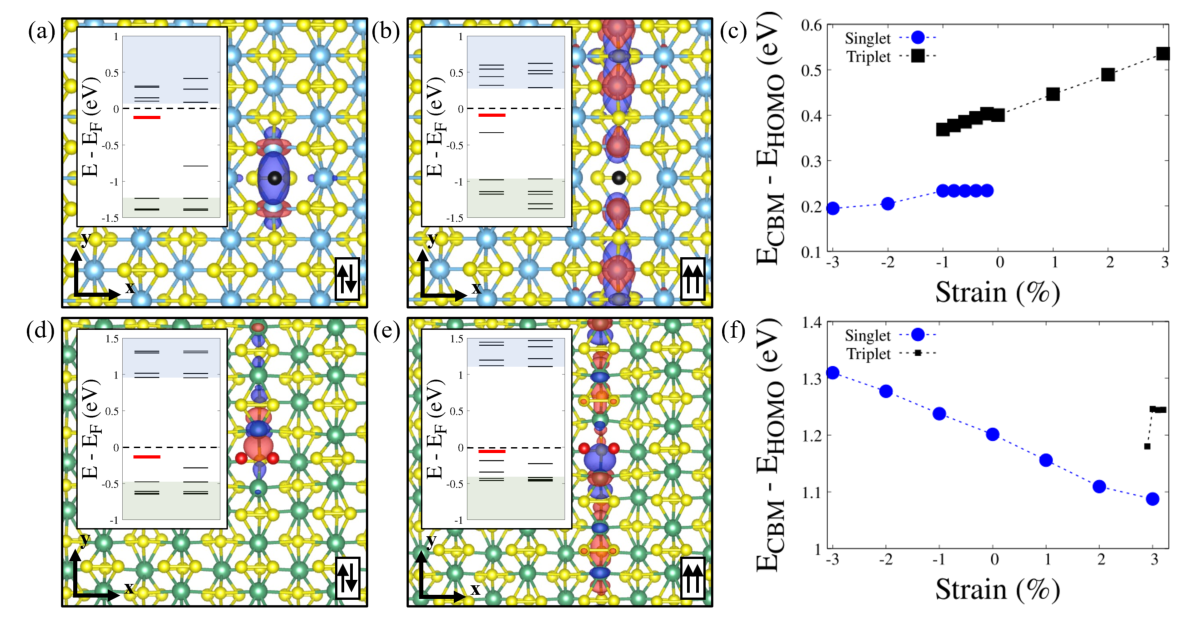}
    \caption{Strain tunable spin states of V$_S$ in TiS$_3$ and V$_D$ in NbS$_3$. (a) Spin singlet and (b) spin triplet state wavefunctions of V$_S$ in TiS$_3$ at $\varepsilon$$_y$ = -0.4 \% and (c) defect position of the HOMO level relative to the CBM. (d) Spin singlet and (e) spin triplet state wavefunctions of V$_D$ in NbS$_3$ at $\varepsilon$$_y$ = 2.9 \%  and (f) defect position of the HOMO level relative to the CBM. The insets of (a), (b), (d), and (f) correspond to the defect electronic structures, following the same conventions described in \ref{e-_structures}. The HOMO levels that are plotted in (c) and (f) are indicated with red lines.}
    \label{defect_wavefunctions}
\end{figure*}

\noindent \textbf{Strain tunable defect spin states.} Similar to bulk TMTC properties, intrinsic vacancy defect properties in TiS$_3$ and NbS$_3$ can be modulated via application of an external strain. The V$_S$ defect in TiS$_3$ and the V$_D$ defect in NbS$_3$ were found to adopt strain-dependent structural polymorphs with associated transitions in their ground spin states. The formation energies of V$_S$ in TiS$_3$ and V$_D$ in NbS$_3$ as a function of strain in the $y$-direction are plotted in Figure \ref{defect_strain_formation}; the remaining strain-formation energy plots are plotted in Figure S.4 \cite{supplement}.

We find that the formation energies of both V$_S$ in TiS$_3$ and V$_D$ in NbS$_3$ decrease as the crystal structures are compressed in the $y$-direction, shown in Figures \ref{defect_strain_formation}(a,c). This trend can be attributed to the strain responses of pristine TiS$_3$ and NbS$_3$ crystals. Both pristine materials incur significant increases in ground state energies as they are compressed in the $y$-direction, while their defect systems show smaller increases in their ground state energies with increasing compressive strain. This can be interpreted as vacancy defects alleviating external compressive strains in both TiS$_3$ and NbS$_3$. This relationship between defect formation energy and external strain in low-dimension materials has been reported on previously \cite{Santra2024}.

Our first principles calculations predict that the V$_S$ defect in TiS$_3$ has competing singlet and triplet ground states at low compressive strains from $\varepsilon_y$ = -1 \% to -0.2 \%, as shown in Figures \ref{defect_strain_formation}(a,b). For zero strain, the formation energy of the triplet ground state lies $\sim$ 0.03 eV below that of the singlet, moving to lower energies for increasing tensile strain. At $\varepsilon_y$ = -0.4 \%, the formation energies of the singlet and triplet ground states flip, indicating that the singlet ground state may be induced through moderate compressive strain. 

We have identified a similar transition for the V$_D$ defect in NbS$_3$. Our calculations indicate that the ground state spin configuration of the V$_D$ defect flips from a singlet to a triplet at a high tensile strain of $\varepsilon_y$ = 2.9 \%, as can be seen in Figures \ref{defect_strain_formation}(c,d). Rather than a reversal of the singlet and triplet state formation energies (as is the case for the V$_S$ defect in TiS$_3$), hybrid DFT calculations show that the triplet state of V$_D$ has two stable configurations, one of which is more stable than the singlet ground state at $\varepsilon_y$ = 2.9 \%. The energy difference between the singlet and triplet ground states at zero strain is about 0.18 eV, suggesting that the singlet state is considerably more stable in the absence of an external strain.

The defect wavefunctions of the strain-dependent ground states of V$_S$ in TiS$_3$ and V$_D$ in NbS$_3$ are shown in Figure \ref{defect_wavefunctions} as well as the strain dependence of HOMO levels relative to the CBM. The singlet wavefunctions are well isolated to the nanowire growth axis but show a large dispersion in that direction in the case of the spin triplet states (Figure \ref{defect_wavefunctions}(a,d)). Figure \ref{defect_wavefunctions}(c) shows how the position of the HOMO level of both the singlet and triplet state relative to the CBM in TiS$_3$ increases as tensile strain is applied, consistent with the relationship established between $\varepsilon_y$ and the electronic band gap. The electronic band gap of TiS$_3$ closes by approximately 0.1 eV at $\varepsilon_y$ = -1 \%; at the same compressive strain, the HOMO level is found to lie about 0.03 eV closer to the CBM relative to the uncompressed case. Thus, the defect comprises a somewhat shallow in-gap state that can hybridize with states at the CBM. The spin singlet state is favorable for $\varepsilon_y \leq$ 0.4 \% and has a ZPL of 0.26 eV.

The position of the spin triplet HOMO level relative to the CBM in NbS$_3$ does not vary significantly at $\varepsilon_y \geq$ 3 \% (Figure \ref{defect_wavefunctions}(f)). However, the position of the spin singlet HOMO level has a span of about 0.21 eV across the range of compressive and tensile strains. In this same range, the band gap of NbS$_3$ varies from 1.03 to 1.41 eV. These findings suggest that functionalization of spin-photon entangled qubits will require additional deformation to better isolate defect systems within the material band gaps \cite{Parto2021}. The high-spin state achieved at $\varepsilon_y \geq$ 2.9 \% exhibits a ZPL of 0.83 eV (1500 nm). 

Examination of defect wavefunctions in Figure \ref{defect_wavefunctions}(a, b, d, e) reveals that quasi-1D geometries of TMTC monolayers act to confine defect states to a uniaxial spread. Specifically, the anisotropic bonding well localizes defect states to the $y$-direction, i.e., the predominately covalent bonding direction, similar to the hybridization of near-band edge states \cite{Kang2016}. Unlike defects in 2D vdW materials that exhibit biaxial localization \cite{Ping2021,Zeng2024}, confinement of defect states to a single direction in TMTCs has two distinct advantages. The defect states are well localized along a single direction, which is ideal for QIS applications where indistinguishable single-photon emission is necessary for scalability. On the other hand, hybridization of defect states with bulk states make their properties sensitive to external stimuli (such as strains or twists \cite{Sutter2023}) that affect their electronic structures. Strategies to improve confinement of defect states in TMDCs are known to be necessary for overcoming hybridization with bulk states \cite{Parto2021}, so similar measures will likely be needed to tailor spectral properties of TMTC defects toward single-photon emission.

Our remaining strain tuning calculations(Figure S.4 \cite{supplement}), including V$_D$ in TiS$_3$ and V$_S$ in NbS$_3$ at varied values of $\varepsilon_x$ and $\varepsilon_y$, did not show transitions in their ground spin states. We previously discussed how strain in the $x$-direction has a more significant impact than strain in the $y$-direction on the electronic structures of TMTC nanowires (Fig. S.1 \cite{supplement}). We similarly expect that the spin and optical properties of TMTC point defects will be more strongly perturbed by $y$-axis strain. Our calculations spotlight intrinsic vacancy defects in quasi-1D TiS$_3$ and NbS$_3$ nanowires as an attractive material platform for semiconductor qubits with tunable optical emission wavelengths and spin state toggling.

\section{Conclusion.}

In summary, we present an \textit{ab initio} analysis of the interplay between external, anisotropic strain and the physical behaviors of vacancy defects in quasi-1D TiS$_3$ and NbS$_3$ nanowires. We demonstrate that the anisotropic structures of TMTC nanowires make them interesting candidate materials for hosting spin-photon qubits. This anisotropy results in axis-dependent material responses to strain; the weak covalent bonding in the $x$-direction also acts to confine defect wavefunctions along the $y$-axis. By applying compressive and tensile strains in the $y$-direction, the band gaps of TiS$_3$ and Nb$_3$ are tunable by $\pm$ 0.23 eV and $\pm$ 0.21 eV, respectively. Additionally, we show that the V$_S$ defect in TiS$_3$ and the V$_D$ defect in NbS$_3$ can adopt competing defect configurations associated with spin state transitions. A modest compressive strain of $\varepsilon_y$ = -0.4 \% can shift the formation energy of V$_S$ in TiS$_3$ such that the singlet ground state becomes energetically preferred to its triplet ground state. This electronic rearrangement also exhibits a small decrease in the ZPL, with a drop from 0.31 eV to 0.26 eV. On the other hand, a large tensile strain of $\varepsilon_y$ = 2.9 \% was found to induce a triplet ground state in the V$_D$ defect in NbS$_3$, with an increase the in ZPL from 0.65 eV to 0.83 eV (1500 nm). 

We note that the strains required to modulate the defect spin states (-0.4 \% to 2.9 \%) are attainable experimentally \cite{Liu2023,Yang2021}. It was shown that 1D vdW materials, such as NbSe$_3$ and TiS$_3$, exhibit an enhanced Young's modulus (6-8 fold increase compared to the bulk modulus) when the nanowire diameter is below a certain threshold ($\sim$ 50-80 nm). This is in line with our recent experimental work \cite{Pham2024} in which several TiS$_3$ and NbS$_3$ nanowires synthesized by chemical vapor deposition, have a diameter less than 50 nm and support a high degree of bending (Figure \ref{Pristine}(c)). Regarding the defect formation, as we discussed, S vacancies have been reported as very common defects in metal chalcogenides because of their high vapor pressure compared to those in transition metals. In the future work, in addition to introducing S defects via a bottom-up synthesis, we will explore energetic beam irradiation as a deterministic pathway to create S vacancies on demand. Our work paves the way to introduce 1D vdW materials, nanowires of transition metal trichalcogenides, as a tunable platform to host defect-induced spin defects for quantum communication and sensing.

\noindent \textbf{Supporting Information.} Additional details on the first principles calculations can be found in the Supplementary Material. 

\section{Acknowledgments}
\begin{acknowledgments}
The authors acknowledge Advanced Research Computing at Virginia Tech (arc.vt.edu) for providing computational resources and technical support, and NCFL for shared facilities. VI acknowledges support from the National Science Foundation Growing Convergence Research Award 2428507.
\end{acknowledgments}

\bibliographystyle{apsrev}
\bibliography{references}


\end{document}


\title{\Large
Supporting information: \textit{Ab initio} prediction of strain-tunable spin defects in quasi-1D TiS$_3$ and NbS$_3$ nanowires}
\author[1]{Jordan Chapman}
\author[2]{Arindom Nag}
\author[2]{Thang Pham}
\author[1,3,4]{Vsevolod Ivanov}

\affil[1]{Virginia Tech National Security Institute, Blacksburg, Virginia 24060, USA}
\affil[2]{Departing of Materials Science and Engineering, Blacksburg, Virginia 24061, USA}
\affil[3]{Department of Physics, Virginia Tech, Blacksburg, Virginia 24061, USA}
\affil[4]{Virginia Tech Center for Quantum Information Science and Engineering, Blacksburg, Virginia 24061, USA}

\date{}
\maketitle

\noindent\ \textbf{Corresponding author:} vivanov@vt.edu

\begin{itemize}
    \item Figure S.1: Pristine TMTC nanowire response to in-plane strain.
    \item Figure S.2: Visualization of VBM and CBM wavefunctions.
    \item Figure S.3: Electronic band structures of TMTC nanowires at varied in-plane strains.
    \item Figure S.4: Formation energies of the sulfide vacancy (V$_S$) and disulfide vacancy (V$_D$) defects in TiS$_3$ and NbS$_3$.
\end{itemize}

\pagebreak

\begin{figure*}[h!]
    \includegraphics[width=1.0\textwidth]{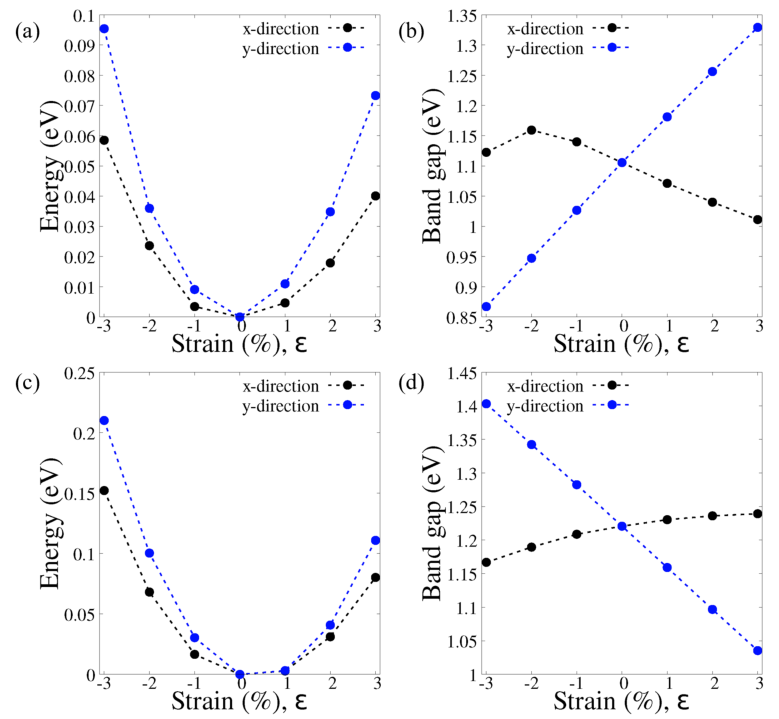}
    \caption{Pristine TMTC nanowire responses to in-plane strain. (a) Formation energy and (b) electronic band gap as a function of strain in the x- and y-directions for TiS$_3$. (c) Formation energy and (d) electronic band gap as a function of strain in the x- and y-directions for NbS$_3$.}
    \label{S1}
\end{figure*}

\pagebreak

\begin{figure*}[h!]
    \includegraphics[width=1.0\textwidth]{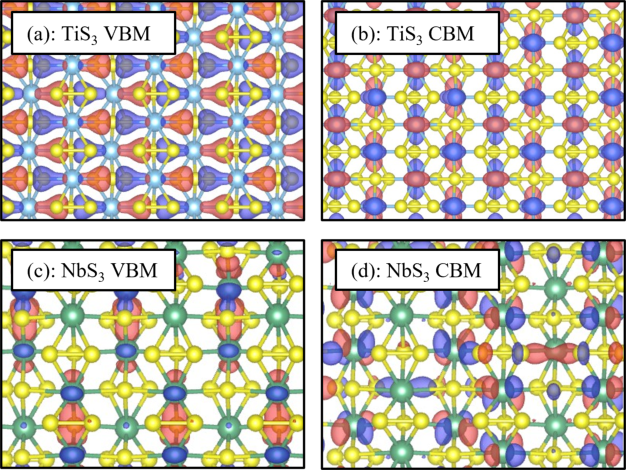}
    \caption{Visualization of wavefunctions population the valence band maximum (VBM) and conduction band minimum (CBM) of TiS$_3$ and NbS$_3$.}
    \label{S2}
\end{figure*}

\pagebreak

\begin{figure*}[h!]
    \includegraphics[width=1.0\textwidth]{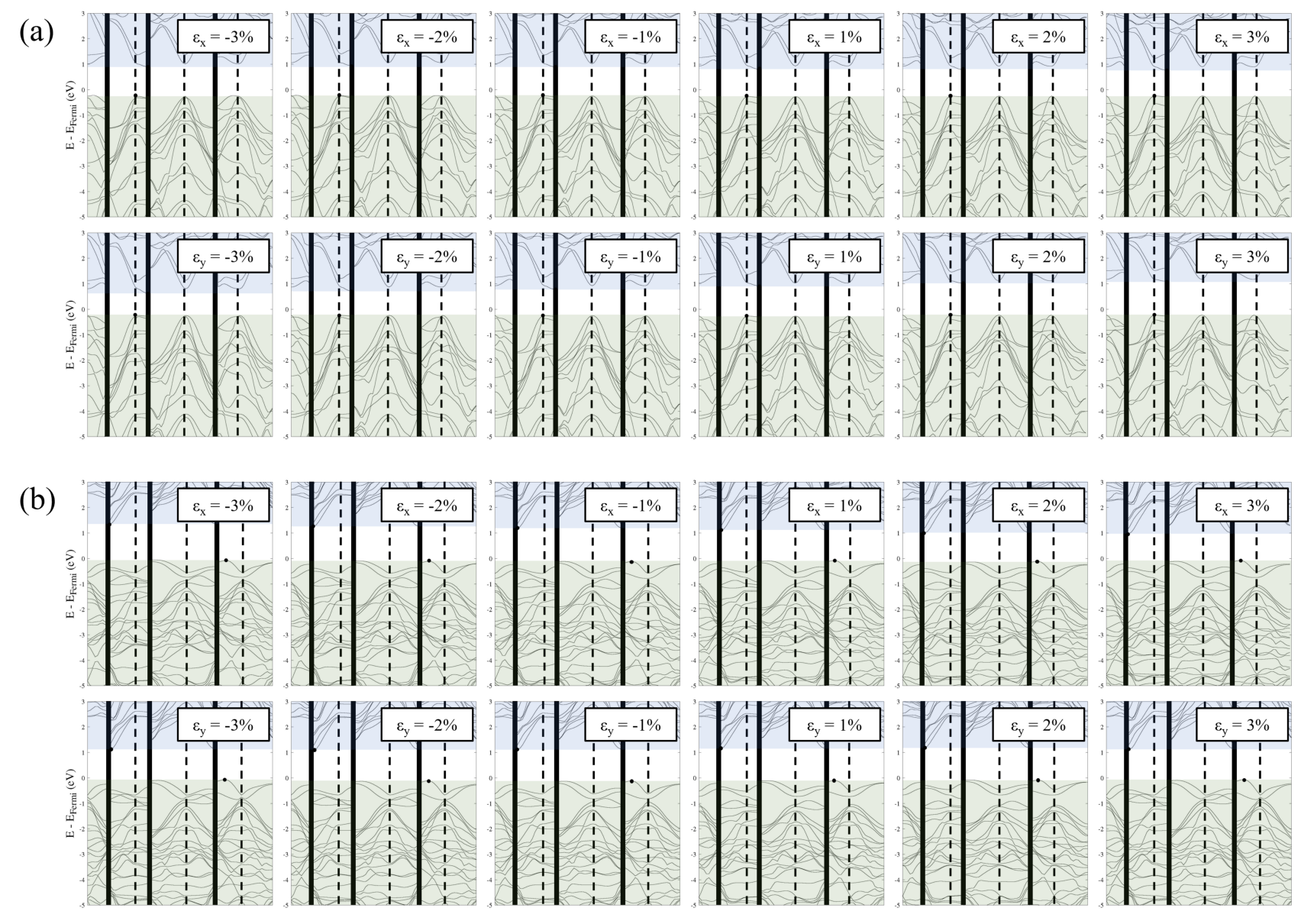}
    \caption{Electronic band structures of (a) TiS$_3$ and (b) NbS$_3$ at varied in-plane strains.}
    \label{S3}
\end{figure*}

\pagebreak

\begin{figure*}[h!]
    \includegraphics[width=1.0\textwidth]{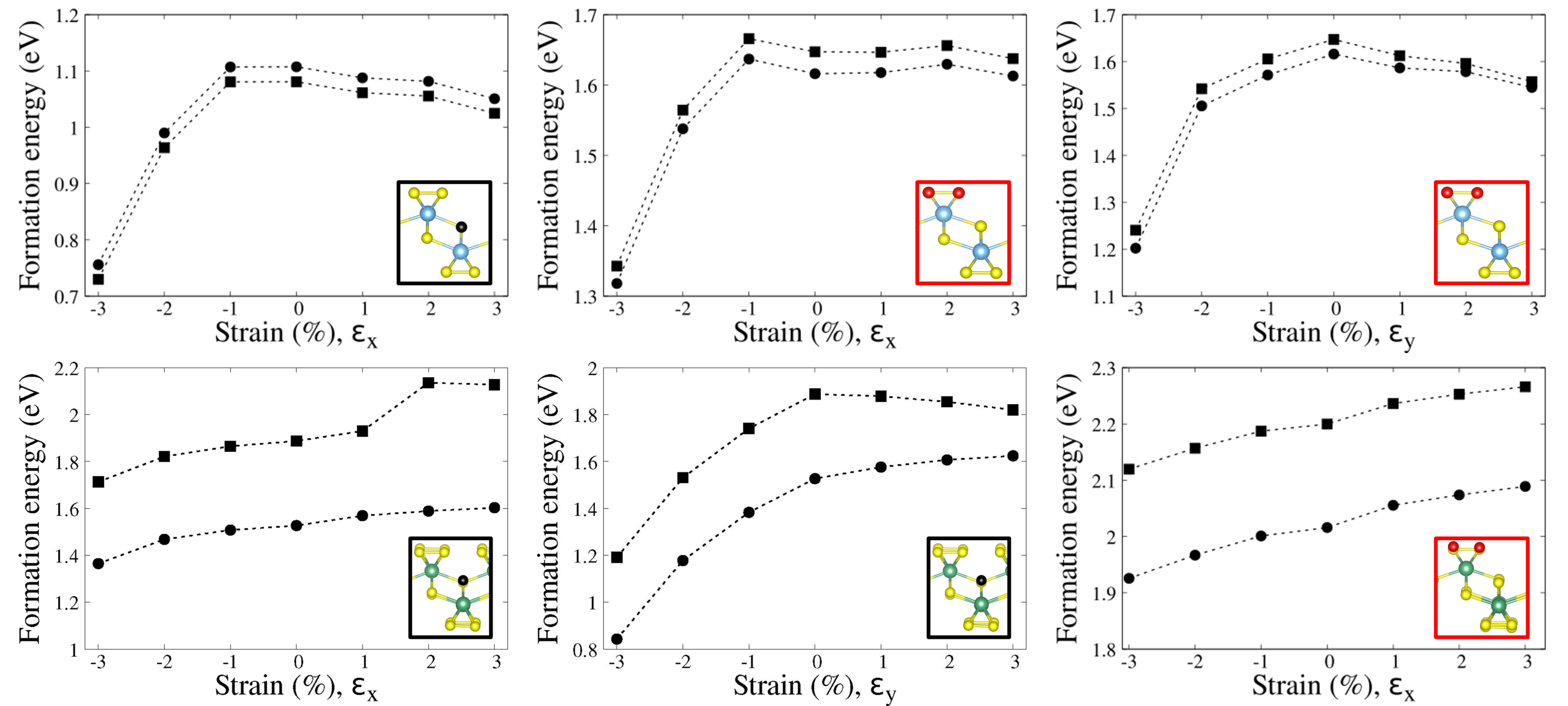}
    \caption{Formation energies of the sulfide vacancy (V$_S$) and disulfide vacancy (V$_D$) defects in TiS$_3$ and NbS$_3$.}
    \label{S4}
\end{figure*}